\documentclass[journal]{IEEEtran}
\ifCLASSINFOpdf
\else
\fi
\usepackage{amsmath}
\usepackage{bbm}
\usepackage{array}
\usepackage{url}

\usepackage[numbers]{natbib} 
\usepackage{graphicx,subcaption}
\usepackage{booktabs}

\begin{document}

\newcommand{\dhl}[1]{\textcolor{black}{#1}}
%
\title{Covariance Estimation and its Application in Large-Scale Online Controlled Experiments}
%
%
%

\author{Tao~Xiong, Yihan~Bao, Penglei~Zhao, and~Yong~Wang} 
\maketitle
\begin{sloppypar}

\begin{abstract}
During the last few decades, online controlled experiments (also known as A/B tests) have been adopted as a golden standard for measuring business improvements in industry. In our company, there are more than a billion users participating in thousands of experiments simultaneously, and with statistical inference and estimations conducted to thousands of online metrics in those experiments routinely, computational costs would become a large concern. In this paper we propose a novel algorithm for estimating the covariance of online metrics, which introduces more flexibility to the trade-off between computational costs and precision in covariance estimation. This covariance estimation method reduces computational cost of metric calculation in large-scale setting, which facilitates further application in both online controlled experiments and adaptive experiments scenarios like variance reduction, continuous monitoring, Bayesian optimization, etc., and it can be easily implemented in engineering practice.
\end{abstract}


%
\IEEEpeerreviewmaketitle

\section{Introduction}
%
%
%
%
Over the past few decades, online controlled experiments are proven to embody the best scientific design for establishing the casual relationship between treatment effect and users' observable behaviour \cite{kohavi2009controlled}. Although the statistical theory underlying online controlled experiments is well developed \cite{manzi2012uncontrolled}, there are difficulties in applying these methods to large-scale systems. In 2010, Google  \cite{tang2010overlapping} first proposed an overlapping experiment infrastructure to tackle these problems, and later the idea of overlapping infrastructure is adopted by many giant companies, e.g., Facebook  \cite{bakshy2014designing}, LinkedIn  \cite{xu2015infrastructure}, Twitter  \cite{al2011experimental}, and also our company.

Given the intrinsic complexity of overlapping infrastructures, there are two significant difficulties in applying statistical inferences, one is how to rigorously define ``metric'' and ``treatment effect'' in this setting; the other, regarded as a top challenge for many companies, is how to compute the metrics at scale reliably and efficiently  \cite{gupta2019top}. We first introduce the basic concepts of overlapping experiment infrastructure and the classical Rubin casual model  \cite{angrist1996identification,rubin1974estimating}. Under the framework of Rubin causal model, we define the metrics in online controlled experiments mathematically in order to better adapt to complex real world situations and facilitate further analysis. Then, if we want to evaluate multiple metrics, assessing the correlation within them is a major concern. However, calculating the covariance among large-sample groups could be computationally unacceptable  \cite{fan2016overview}. In order to conquer this problem, we propose a covariance estimation method, which helps us to 
reduce computational costs, so that it is applicable in large-scale settings. Theoretical induction and numerical simulation are both carried out to evaluate our covariance estimation method in terms of accuracy and computational efficiency. 

Moreover, our covariance estimation method can also be applied to other tasks in online metric analytics which rely on the estimation of covariance and variance (variance can be considered as a special form of covariance). We proposed 3 applications: variance reduction  \cite{deng2013improving, guo2015flexible,xie2016improving}, continuous monitoring  \cite{deng2016continuous} and Bayesian optimization  \cite{letham2019constrained}, and we evaluate our covariance estimation with numerical examples accordingly. Besides, the benchmarks of our proposed method deployed in our experimentation system are also illustrated.

The main contributions of this paper are summarized as follows:

\begin{itemize}
    \item We give a rigorous definition of online metrics under the framework of Rubin causal model, which can integrate classical causal inference theories with modern large-scale experimental design.
    \item We propose a novel and computational efficient method for estimating covariance with more flexibility to trade off between computational costs and precision in online metric analytics. 
    \item We demonstrate that our covariance estimation method can also be applied to other tasks in A/B testing platform. To our knowledge, no other companies have adopted such an ameliorated covariance estimation algorithm in engineering practice.
\end{itemize} 
For reproducibility of all the simulations in the paper, we provide all related code under a public Github repository at \url{https://github.com/xt2357/covariancesimulation}

\section{Background}
\subsection{Overlapping Experiment Infrastructure}
We first introduce the overlapping infrastructure in order to better illustrate this particular application, although the proposed method itself does not rely on the infrastructure. 

\subsubsection{Basic Concepts}
In the context of A/B Testing platform, for every experiment we have live traffic as incoming data, and traffic is segmented to different experiment groups. Typically, user is the randomization unit, where users are distributed to groups with different treatments to study their user-level metrics in randomized experiments  \cite{deng2011choice}.

To further explain the overlapping infrastructure, we introduce 3 key concepts below:
\begin{itemize}
    \item {A \textbf{Domain}} is a segmentation of incoming live traffic. 
    \item {A \textbf{Layer}} is a partition of system parameters, where each subset of parameters is carefully determined to ensure that the parameters in different layers have little interaction. Experiments can be carried out within a layer. 
    \item {An \textbf{Experiment}} is a segmentation of traffic where zero or more system parameters in the current layer can be assigned alternate values. 
\end{itemize}


In overlapping infrastructure, layers and domains can be nested. Domains contain layers, layers contain experiments, but can also contain domains; this nesting framework introduces more flexibility to partition the system parameters. 

\subsubsection{Traffic Diversion Algorithm}
Domains, layers and experiments are organized into a tree-like structure, as illustrated in Figure 1. 

\begin{figure}[h]
  \centering
  \includegraphics[width=\linewidth]{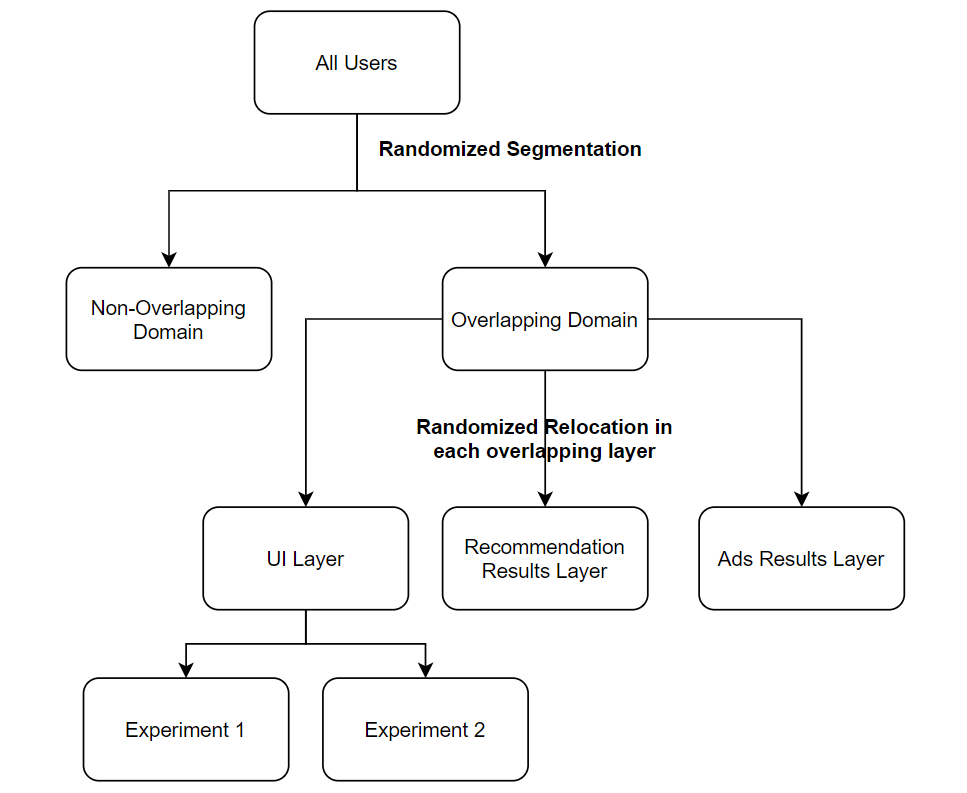}
  \caption{A typical tree structure of domains and layers in overlapping infrastructure.}
\end{figure}

With this tree structure, the incoming traffic is diverted by the randomization unit, for example, user is the randomization unit in the our A/B Testing platform, and every request is diverted by its unique user-id. In the example of Figure 1, every user is randomly diverted to only one domain, but those users in the overlapping domain are in these 3 layers simultaneously. A randomized reallocation mechanism of users is applied to every layer of the overlapping domain to ensure that users are independently diverted, that is to say, 
every user in the overlapping domain is independently re-randomized into the experiments within each layer, and is re-randomized into groups within the allocated experiment. 


The randomization mechanism described above can be achieved using a hash function $f$ with a seed: for example, given the user-id of a user, we calculate $f$(user-id, seed) modulo 10000, then all users with the same mod will be grouped together for traffic diversion. Assume we have a set of users with mod equals to 345, then they will all be diverted into the overlapping domain if mod 345 is configured to be included in this domain as a stream of traffic. With a unique seed for each randomization, we can assume that all the randomization processes within domains, layers and experiments are independent. For example, the seed used in the segmentation of the non-overlapping domain and the overlapping domain will not equal to the seed used in the randomized relocation of the UI layer. In fact, the diversion algorithm described above defines a  \textit{deterministic assignment} of all potential users into domains, layers and experiments.

\subsection{Rubin Causal Model}
\label{sec:rubin-model}

Our covariance estimation method is built on the framework of Rubin casual model  \cite{angrist1996identification,rubin1974estimating,sekhon2008nrm}. A key notion underlying the Rubin causal model is \textit{potential outcomes}. For example, a particular unit $i$ would have an outcome of $Y_1(i)$ if it is exposed the treatment ($T=1$), and would have an outcome of $Y_0(i)$ if it is exposed the treatment ($T=0$); potential outcome is independent of whether it actually receives treatment or not. To measure  \textit{unit-level casual effect}, we should calculate the difference of corresponding outcomes with or without treatment, $Y_1(i) - Y_0(i)$; however, it is impossible to observe both the potential outcomes with or without treatment at the same time, and this dilemma is the ``fundamental problem of causal inference''  \cite{rubin1974estimating,holland1986statistics}. 

Although we cannot directly measure the unit-level treatment effect due to this dilemma, we can still estimate the average treatment effect (ATE) over the entire population in randomized experiments by calculating the difference of means between the observations from treatment groups and the observations from control groups, which is an unbiased estimator of ATE. \cite{rubin1974estimating}.

Noted that there is an important assumption throughout the above procedure, called ``stable unit treatment value assumption'' (SUTVA)  \cite{rubin1986comment}, which requires that ``the observation of potential outcome on one unit should not be affected by the particular assignment of treatments to the other units'', and this is why we can assume all units in the treatment group are only observed with $Y_1's$.

\section{Methodology}
\label{sec:methodology}
Under the framework of Rubin causal model, we first give the rigorous definition of online metrics, then introduce the covariance estimation problem of online metrics.
\subsection{Definition of Online Metrics}
\label{sec:online-metrics}

Suppose for a randomized experiment, there are experiment groups $1,2, ... G$, users $1,2, ..., U$, and metrics $1,2, ..., M$, where $G$, $U$ and $M$ are predetermined nonrandom quantity. Then for every pair of experiment group $g$, user $u$ and metric $m$, we define $Y(g,u,m)$ to be the corresponding potential outcome, and $I_g(u)$ to be the indicator of whether user $u$ is actually assigned to the group $g$, with $I_g(u)=1$ meaning user $u$ is actually in group $g$. Another indicator $Z(g,u,m)$ denotes whether this potential outcome $Y(g,u,m)$ is successfully observed in actual, with value 1 indicating success. Then the sum of all observed outcome for metric $m$ within the group $g$ is defined as $S(g,m) = \sum_{u \in \{1,2, ...U\}}{I_g(u)Y(g,u,m)Z(g,u,m)}$, and the average value of metric $m$ for the group $g$ is $A(g,m)= \sum_{u \in \{1,2, ...U\}}{I_g(u)Y(g,u,m)Z(g,u,m)}/$ $\sum_{u \in \{1,2, ...U\}}{I_g(u)Z(g,u,m)}$

More generally, the calculation of most complex metrics can be derived by some function $f$ mapping from the average metrics. We denote a complex metric for a given group $g$ as $V_{g}$, where $V_{g} = f(A(g,1), A(g,2), ...,A(g,M))$. Moreover, in the setting of online controlled experiments, for a given group $g$, $I_g(u)$ of every user $u \in \{1,2, ...U\}$ are considered to be i.i.d. random variables. Thus, for any group $g$, the average metrics in vector form $\vec{A}(g)=(A(g,1), A(g,2), ...,A(g,M))$ would be asymptotically (multivariate) normal distributed in the "big data" setting \cite{van2000asymptotic, deng2018applying}:
$$(\vec{A}(g)-\vec{\alpha}_g) \xrightarrow{D} \mathcal{N} (\vec{0},\Sigma_g)$$ where $\vec{\alpha}_g$ is the true value $\vec{A}(g)$ converges to, and $\Sigma_g$ is a symmetric positive semi-definite covariance matrix for average metrics of group $g$.
Then by Delta method \cite{deng2018applying}, the complex metric $V_{g}$ is also asymptotically normal distributed when $f$ is differentiable:
 $$f(\vec{A}(g)) \approx  f(\vec{\alpha}_g)+\nabla f(\vec{\alpha}_g )^{T}\cdot (\vec{A}(g)-\vec{\alpha}_g )$$ which implies that $$\left(f(\vec{A}(g))-f(\vec{\alpha}_g )\right)\,{\xrightarrow  {D}}\, \mathcal{N} \left(0,\nabla f(\vec{\alpha}_g )^{T}\cdot \Sigma \cdot \nabla f(\vec{\alpha}_g )\right)$$
For simplicity, we will use the average metric for illustration in the rest of the paper, but the conclusions can be easily generalized to the other complex metrics using differentiable functions.

\subsection{The Covariance Estimation Problem}
\label{sec:cov-estimation-problem}
Consider we are doing an online experiment, and we are interested to discover the relationships between some key metrics over some specific time period. For example, in an experiment of social media data analysis, we would like to check if the average number of exposures to a video leads to a higher average number of comments for this video in the last week, and we can formalize these kind of experiments as following:

In an experiment, the average of metric $m$ for group $g$ during period $t$ is defined as 
$$ A_{t}(g,m)= \sum_{u}{I_g(u) Y_{t}(g,u,m) Z_{t}(g,u,m)} / $$ 
$$ \sum_{u}{I_g(u) Z_{t}(g,u,m)},$$
and we denote the numerator as $S_t(g,m)$, and denominator as $N_t(g,m)$.

After this rigorous definition based on potential outcome,  we know that the value of actual observation of metric $m$ for user $u$ in group $g$ over period $t$, denoted as $X_{t}(g,u,m)$, would simply be equivalent to $Y_{t}(g,u,m)$, given that $I_g(u) = 1$ and $Z_{t}(g,u,m) = 1$.


Now the problem is, in order to measure the relationship of metric $m_1$ and $m_2$, how to measure their covariance efficiently?

\subsubsection{Existing Methods}
In large-scale experimentation platform, observations from different users are often considered i.i.d. samples since the population is very large and the size of the observations is small compared to the size of the population. In the case of estimating the covariance of two metrics in the same period $t$, suppose we have observed two metrics value $m_1$ and $m_2$ for each user $u$, then estimating the covariance of two sample averages is very similar to estimating the sample covariance, which is trivial. 

For estimating the covariance of two periods $t$ and $t'$ of the same metric, if for every user we observe both values in the two periods $t$ and $t'$ successfully, the covariance can also be estimated using the sample covariance formula; but in most real-world cases, not all users use our product in both periods $t$ and $t'$ , and it will cause the problem of \textit{missing data}.

The naive approach is to calculate the sample covariance only with the observations of users who appear in both periods, but it only works when the data is missing completely at random  \cite{deng2018applying}, which is not a reasonable assumption in reality, because in fact inactive users would systematically suffer from more missing data than active users.

There are many existing works on handling missing data, one approach is based on modelling the missing data mechanism, which requires rather strong assumptions on moments of squares and crossproducts of elements \cite{leon2003semiparametric}. Another approach adopts the Delta method after \textit{data augmentation}, which requires calculating the user-level sample covariance with the augmented data \cite{deng2018applying}. 

We notice that all these existing methods of covariance estimation are  based on user-level data aggregation and manipulation. However, in the context of large-scale online controlled experiments, where we have thousands of groups, metrics, and tens of millions of users involved in different experiments with periods of several weeks, the user-level data  processing are computationally expensive. Imaging a 30-days experiment that contains ten million users using our company's product every day, and continuously we collect their daily activity data under different metrics, then even a single covariance estimation process for only one metric between two days needs a user-level data join of $10,000,000$ observations; and this process will repeat $30\times 30/2=450$ times for estimating pair-wise covariance among 30 days, which is a huge requirement for computational resources in real world applications.

\subsubsection{Our Bucket-Based Method}
We present a novel idea for estimating the covariance with much lower computational costs and more flexibility, and solving the data missing problem without the need of modelling the data missing mechanism. The method we present is based on a preprocessing procedure, which is called ``bucketing". 

In the realization of ``bucketing", we introduce an extra deterministic randomization process for assigning all users into different buckets $1,2, ... B$. Noted that this randomization process should be independent with the traffic diversion process described in section 2 by choosing a totally different hash seed: given any user-id, we can calculate the hash value of the user-id, and get the remainder $b$ of its hash value modulo $B$, then users with the same remainder $b$ are assigned into the bucket $b \in \{1,2, ..., B\}$ respectively. Formally, we use indicator $I_{b}(u)$ to represent the bucketing result of the user $u$, $I_{b}(u)=1$ if and only if the user $u$ is assigned to bucket $b$, otherwise 0. 

With this bucketing algorithm, we can reduce the observations into a number of $B$ potential outcome sums, for example, given $n$ observations of the metric $m$ for a group $g$ in period $t$, we sum up the observations within each bucket respectively, and the potential outcome sum for bucket $b$ with respect to group $g$ and metric $m$ within period $t$ is denoted as $S_{t}(g,m,b)=\sum_{u}{I_g(u) I_{b}(u) Y_{t}(g,u,m) Z_{t}(g,u,m)}$, it also equals to the summation of those successful observations from users who are assigned to bucket $b$. Given assumptions from Rubin causal model and the independence property of the hash function used in traffic diversion and bucketing, we prove in the following Theorem 1 that this estimator is applicable in calculating covariance between any two average metrics, and for every average metric it can be calculated with only $B$ potential outcome sums, which significantly reduces computational costs when an appropriate $B$ is chosen.

For any potential outcome sums of metrics $m$ and $m'$ across all buckets for group $g$ at different period $t$ and $t'$ respectively, denoted as $S_{t}(g,m)$ and $S_{t'}(g,m')$, we first define the bucket-level sample covariance as $$K(S_{t}(g,m),S_{t'}(g,m'))$$ 
which equals to 
$$
\begin{aligned}
\frac{1}{B-1} \sum^B_{b=1} &{(S_{t}(g,m,b)-\bar{S}_{t}(g,m,b))} \times \\
&{(S_{t'}(g,m',b)-\bar{S}_{t'}(g,m',b))},
\end{aligned}
$$
where $\bar{S}_{t}(g,m,b)$ is the average of corresponding values across $B$ buckets.  
\newline
\newline
\textbf{Theorem 1.} 
\newline
For any period of $t$ and $t'$ for metrics $m$ and $m'$ of group $g$, the corresponding average metric is $A_{t}(g,m) = {S_{t}(g,m)}/{N_{t}(g,m)}$, and $A_{t'}(g,m') = {S_{t'}(g,m')}/{N_{t'}(g,m')}$, and for simplicity we denote them as $A_{t}$, $A_{t'}$, $S_{t}$, $S_{t'}$, $N_{t}$, $N_{t'}$, then the estimation for its covariance $\text{cov}[A_{t}(g,m),A_{t'}(g,m')]$:
 $$
 \begin{aligned}
 \text{cov}[A_{t},A_{t'}] = & CB[\frac{K(S_t,S_{t'})}{N_t N_{t'}}+\frac{S_t S_{t'} K(N_t,N_{t'})}{N_t^2 N_{t'}^2} \\
 &-\frac{S_{t'}K(S_{t},N_{t'})}{N_{t}N_{t'}^2}-\frac{S_{t}K(S_{t'},N_t)}{N_t^2 N_{t'}}]
 \end{aligned}
 $$ 
where $B$ is the number of buckets and $C$ is the correction term satisfying
$$C=1-E[I_g(u)].$$
$E[I_g(u)]$ is the expectation of $I_g(u)$ across any user $u$. 
\newline
Noted it does not matter which user to choose here, because $I_{g}(u)$ of different users are i.i.d. random variables. In fact, $E[I_g(u)]$ is the probability of assigning users to group $g$, and the correction term can be approximating to 1 when the size of observations is very small compared to the size of population. The proof of Theorem 1 is in the appendix.

\subsection{Numerical Examples and Interpretation}
\label{sec:num-examples}
\subsubsection{Settings}
First, we simulate $N =  10,000$ users as a population using the following steps: \begin{itemize}
\item Generate $Y_{t}(g,u,m)$ and $Y_{t'}(g,u,m')$ from $\mathcal{N}(\vec{\mu}, \Sigma)$ for each user, and the covariance matrix $\Sigma$ is constructed such that $Y_{t}(g,u,m)$ and $Y_{t'}(g,u,m')$ are correlated.
\item For a metric $m$ in observation period $t$, define the activeness of user $u$ as the percentage of its observation value among all users in group $g$. Formally, for any $Y_{t}(g,u,m)$, we define corresponding activeness $e_{t}(g,u,m)=\sum_{u'}\mathbbm{1}\{u'|Y_{t}(g,u',m)<Y_{t}(g,u,m)\}/N$, and the same for $Y_{t'}(g,u,m')$.
\item Generate $Z_{t}(g,u,m)$ from user activeness for probabilistic modeling of missing data, specifically, for each user $u$, there is a probability of $\text{min}(0.5, 1-e_{t}(g,u,m))$  for the corresponding observation of $Y_{t}(g,u,m)$ to be missing, thus we can generate $Z_{t}(g,u,m)\sim \textit{Bern}(1.0-\text{min}(0.5, 1-e_{t}(g,u,m))),$ and the same for $Z_{t'}(g,u,m')$.
\end{itemize}

Then we simulate the following procedure 100,000 times: each time we generate $I_{g}(u) \sim \textit{Bern}(0.1)$ for each user in the population, and get the corresponding observations by $Y_t(g,u,m)$, $Y_{t'}(g,u,m')$, $Z_{t}(g,u,m)$ and $Z_{t'}(g,u,m')$ for users with $I_g(u) = 1$. A bucketing process is then carried out, assigning each user to one of the buckets $1,2, ... B$ independently, and with the same probability of $1/B$. 

For comparison, we run the naive approach, the data augmentation approach and our bucket-based approach for estimating covariance; the true covariance can be estimated by the sample covariance calculated from $A_{t}(g,u,m)$ and $A_{t'}(g,u,m')$ among 100,000 trials, and the mean and standard deviation of the 100,000 estimations by each approach is compared to the true covariance for evaluating the estimation. 

\subsubsection{Results and Interpretation}

\begin{table}[h]
  \caption{Simulation Results: Ground truth, average, standard deviation and the time for calculation of covariance in millisecond.}
  \label{tab:sim-results}
  \centering
  \begin{tabular}{lrrrr}
    \toprule
    Method&Ground Truth&Avg.&SD.&Time\\
    \midrule
    Naive& 8.825& 21.085& 1.708&135\\
    Data Augmentation& 8.825& 9.790& 0.685&1219\\
    Bucketing of $B=100$& 8.825& 8.792& 1.684&155\\
    Bucketing of $B=200$& 8.825& 8.807& 1.269&298\\
    Bucketing of $B=500$& 8.825& 8.801& 0.931&733\\
    Bucketing of $B=1000$& 8.825& 8.802& 0.790&1458\\
  \bottomrule
\end{tabular}
\end{table}

 We can observe from Table \ref{tab:sim-results} that with our bucket-based covariance estimator, the average is closer to the ground truth of the covariance compared to other methods; also the fluctuation decreases as the number of buckets increases. However, the computational time also increases as the number of buckets increases, thus we need to trade off between computational costs and accuracy by choosing an appropriate number of buckets.

Moreover, we find that the data augmentation approach that assumes i.i.d. observations is biased upward, which can be explained by the fact that we had sampled 10\% of the population distributed to different groups without replacement, but the observations can be treated as i.i.d. only when the sample size is small compared to the population. To better illustrate this problem, we adopt the same setting as the previous section, and simulate the data augmentation method with different sampling ratios, the results are summarized in Table \ref{tab:sim-data-aug}.

\begin{table}[h]
  \caption{Simulation Results: Data augmentation method with different sampling ratios. The results indicate that the upward bias decreases when the ratio decreases.}
  \label{tab:sim-data-aug}
  \centering
  \normalsize
  \begin{tabular}{lrrr}
    \toprule
    Ratio&Ground Truth&Avg.&SD.\\
    \midrule
    0.2& 3.862& 4.798& 0.222\\
    0.1& 8.639& 9.572& 0.689\\
    0.05& 18.081& 19.065& 1.977\\
    0.01& 94.121& 94.164& 22.596\\
  \bottomrule
\end{tabular}
\end{table}

In real A/B testing platform, we often encounter the situation of large ratio (more than 5\%) experiment groups in real world, in which case the assumption of independence is no longer applicable. This poses a difficulty to evaluate the metrics accurately, but our bucket-based method is not built upon the i.i.d. assumption of observations, thus can still work well in such situations.

\section{Deployment and Simulation Results}
\label{sec:deployment}
In this section, we introduce three applications of our covariance estimation method: variance reduction, continuous monitoring, and Bayesian optimization.

\subsection{Variance Reduction}
\label{sec:variance-red}
\subsubsection{Background} 
One of the major purposes for carrying out online controlled experiments is to measure whether a treatment or strategy given to users would significantly influence their behaviours, and it can be reflected by the change of key performance metrics. This change is defined as ``average treatment effect'' (ATE) in A/B Testing platform.
To illustrate, for a given metric $m$ from control group $g$ and experiment group $g'$ during a specific period $t$, the average metrics $A_{t}(g,m)$ and $A_{t}(g',m)$ are unbiased estimators of their corresponding true averages of potential outcomes from population, and the estimator of ATE can be written in such form: $\Delta_t(m) = A_{t}(g',m) - A_{t}(g,m)$, which is an unbiased estimator of $\delta = \mathbf{E}(\Delta)$.

As discussed before, the challenge with measuring ATE is the ability to detect it when it indeed exists, usually referenced as ``sensitivity''. One way to improve sensitivity is to carry out variance reduction  \cite{deng2013improving,guo2015flexible,xie2016improving}.
\subsubsection{CUPED}
For the purpose of variance reduction, Deng et al. \cite{deng2013improving} proposed a method called CUPED (Controlled-experiment Using Pre-Experiment Data), in which they introduce a control variate $Y$ with a known expectation $\theta$. Define:
$$\hat{\Delta}_t(m) = \Delta_t(m)-\beta(\bar{Y}-\theta),$$
where $\beta$ can be any constant. It can be easily shown that $\hat{\Delta}_t(m)$ is still an unbiased estimator of $\Delta$, regardless of the value of $\beta$.
Moreover, the variance of $\hat{\Delta}_t(m)$ is 
$$
\begin{aligned}
\label{eq:CUPEDvar}
\text{var}(\hat{\Delta}_t(m)) &= \text{var}(\Delta_t(m) -\beta\bar{Y}) \\
&= \text{var}(\Delta_t(m)) + \beta^2 \text{var}(\bar{Y}) - 2\beta \text{cov}(\Delta_t(m),\bar{Y}),
\end{aligned}
$$
\newline
$\text{var}(\hat{\Delta}_t(m))$ reaches its minimum when $\beta = \text{cov}(\Delta_t(m),\bar{Y})$ $/\text{var}(\bar{Y})$. With this optimal $\beta$ we have:
$$\text{var}(\hat{\Delta}_t(m)) = \text{var}(\Delta_t(m))(1-\rho^2),$$ 
where $\rho$ is widely known as the correlation coefficient. This estimator $\hat{\Delta}_t(m)$ with reduced variance while keeping the unbiased property improves the sensitivity of the online metric $m$, and allows for a more precise assessment of the metric value. For example, in a simulation Deng et al. carried out in the same paper \cite{deng2013improving}, for metric ``queries-per-user", using ``queries-per-user in the 1-week pre-experiment period" as control control variate $Y$, variance reduction rate can reach more than 45 percent.

The major problem here is that we need to estimate the optimal value of $\beta$ when conducting the variance reduction procedure, and it is important to point out that estimating $\text{cov}(\Delta_t(m),\bar{Y})$ can be expensive when group size is gigantic. By applying our covariance estimation method, the cost of estimating the covariance would reduce from joining two data sets on user-level (in most cases larger than ten million) to bucket-level, where typically we choose the number of bucket size less than 1000.

\subsubsection{Numerical Examples}
To illustrate the effectiveness of our covariance estimation method in variance reduction case, we simulate the control variate $Y$ with zero mean, and being correlated with the observations. Then, we want to evaluate the precision of the optimal $\beta$ estimated with the proposed method under different bucket numbers and correlation coefficients.

Let sample size $n=10,000$, and for this fixed sample size, we consider the correlation between $\Delta_t(m)$ and $Y$ while $\rho$ = 0.3, 0.5, 0.6, 0.8; and bucket number $B$ =  50, 100, 200, 500, 1000 with a repetition of 1000 times.


\begin{table}[h]
  \caption{Simulation Results: Relative error of the optimal $\beta$ estimated using our method compared to the optimal value of $\beta$ in theory.}
  \label{tab:sim-var-red}
  \centering
  \begin{tabular}{lrrrrrr}
    \toprule
    Relative Error& B=50& B=100& B=200& B=500& B=1000\\
    \midrule
    $\rho$=0.3&0.1708&0.1197&0.0817&0.0534& 0.0410\\
    $\rho$=0.5&0.0866&0.0574&0.0413&0.0278&0.0201\\
    $\rho$=0.6&0.0630&0.0437&0.0301&0.0189& 0.0142\\
    $\rho$=0.8&0.0252&0.0174&0.0127&0.0080& 0.0059\\
  \bottomrule
\end{tabular}
\end{table}

The results are summarized in Table \ref{tab:sim-var-red}, the relative error of the estimated optimal $\beta$ decreases as the number of buckets increases. It is also interesting to notice that the relative error decreases rapidly when the correlation increases, which indicates that the precision of our method would increase when a stronger correlated control variate is chosen. (The optimal $\beta$ approaches to zero when the correlation approaches to zero, in which case even a small random error of the estimation will lead to a large relative error).

\subsection{Continuous Monitoring}
\label{sec:continuous-mon}
\subsubsection{Background} Hypothesis testing is a powerful tool for conducting statistical inference in online controlled experiments, and the null-hypothesis statistical testing (e.g., t-test or z-test) is widely used in A/B tests, where they summarize the test result with a p-value, and reject the null hypothesis $H_0$ when the p-value is less than the significant level $\alpha$. It is guaranteed that the probability under the null hypothesis of making a Type-I error (reject $H_0$ when $H_0$ is true) is less than that pre-determined $\alpha$. 

In a t-test we cannot continuously monitor the result and then early stop the test once a significant signal is detected (p-value less than $\alpha$), otherwise the probability of making Type I error will not be bounded by $\alpha$ anymore. Instead, before conducting a t-test for an online-controlled experiment, we need to determine the sufficient sample size for the t-test in advance, and after the experiment starts, only make the conclusion when the sample size is reached. This procedure is rather inflexible when carrying out online controlled experiments; for example, the users might need to decide a large sample size in advance while the true effect is easy to detect with a much smaller sample, and the opportunity cost of waiting for extra samples can be large in such cases. A test which allows continuous monitoring can help users detect the true effects as quickly as possible, and adjust the sample size dynamically with more flexibility; but continuous monitoring in the t-test will cause a severe inflation of Type-I error, making the test an invalid inference \cite{johari2017peeking}. 

There are many methods proposed to solve the problem of continuous monitoring, one approach is to construct an always-valid p-value despite continuous monitoring from the users \cite{johari2017peeking}, but this requires rather strong assumptions about the collected data. Another approach is based on Bayesian testing  \cite{deng2016continuous}, where given two prior probabilities $P(H_0)$ and $P(H_1)$ for $H_0$ and $H_1$ to be true respectively, we update the posterior odds by $$\frac{P(H_1|Data)}{P(H_0|Data)}=\frac{P(H_1)}{P(H_0)} \times \frac{P(Data|H_1)}{P(Data|H_0)}$$ The last term above is the likelihood ratio of the observed data, which is also known as ``Bayes factor". Deng et al. proved that continuous monitoring based on specific stopping criterion about the posterior odds can control the false discovery rate (FDR, the proportion of false discoveries among the discoveries) at a predetermined level. To be specific, stopping the monitoring and rejecting $H_0$ when the posterior odds is greater than $K$ can guarantee a FDR upper bound of $1/(K+1)$.

We adapted the Bayes factor method into our experimentation system by applying the bucked-based covariance estimation described in this paper: 

Observed data is distributed over many periods, i.e., each day from Day $1$ to Day $d$ $\{D_1, D_2, ..., D_d\}$ can be considered as different period $t$ of the experiment, then in every period we observe $X_{D_1}(g,u,m), X_{D_2}(g,u,m), ..., X_{D_d}(g,u,m)$ for every user $u$ in group $g$ of metric $m$, and the average value of observations across for every period $t$ is $A_t(g,m)=\sum_{u}{X_t(g,u,m)}/{N_t(g,m)}$ for $t \in \{D_1, D_2, ..., D_d\}$, where $N_t(g,m)$ is the number of observations in period $t$ for group $g$ of metric $m$. Then the Bayes factor based on average metrics $A_{D_1}(g,m),A_{D_2}(g,m),...,A_{D_d}(g,m)$ is calculated and users can monitor the Bayes factor continuously by each period, and stop the experiment once the stopping criterion is satisfied.

The difficulty is that there are overlapping users among periods, the observations of metrics from the same user from different periods are inevitably correlated, and we must calculate the likelihood ratio of  \textit{the whole path} of the observations, otherwise we will get an inaccurate likelihood and be unable to control the FDR. Fortunately, $A_t(g,m)$ of different periods can be considered (multivariate) normal distributed by applying central limit theorem, and the covariance matrix of $A_{D_1}(g,m),A_{D_2}(g,m),...,A_{D_d}(g,m)$ can be estimated efficiently by our bucket-based covariance estimator, and the likelihood ratio can be calculated using multivariate normal density formula. 

\subsubsection{Numerical Examples}
Suppose for group $g$ we have periods of $d = 30$ days and $N=4000$ users, and for every user the average metric from Day 1 to Day $d$ is an i.i.d. random vector with a length of $d$ from a multivariate normal distribution where the samples of different periods are correlated, the mean of this normal distribution is  $\vec{\mu}= 0$ under the null hypothesis $H_0$, and $\vec{\mu}= 0.3$ under $H_1$. We simulate 10000 runs under $H_0$ and other 10000 runs under $H_1$, with 20000 runs of the Bayes factor continuous monitoring are simulated in total, and the prior odds of $H_1$ over $H_0$ is 1. 

We can add observations from all users together to get the average metric over all periods for group $g$, $A_{D_1}(g,m),$ $A_{D_2}(g,m),...,A_{D_d}(g,m)$ under both $H_0$ and $H_1$ respectively; then calculate Bayes factor for each period $t$, and reject $H_0$ when the observed Bayes factor is larger than $9$, guaranteeing a FDR bound of $1/(9+1)=0.1$.

Here we implement three methods to derive the likelihood $P(Data|H_0)$ and $P(Data|H_1)$ for calculating the Bayes factor over every period $t$: the non-covariance method is to treat $A_{D_1}(g,m),A_{D_2}(g,m),...,A_{D_d}(g,m)$ as independent random variables, where the density of every $A_t(g,m)$ is calculated respectively by normal density formula (using the true variance of each $A_t(g,m)$) and the likelihood of the whole path of $A_{D_1}(g,m),A_{D_2}(g,m),...,A_{D_d}(g,m)$ is the multiplication of all these densities; the second method is that we calculate the likelihood by the multivariate normal density formula using the true covariance matrix of $A_{D_1}(g,m),A_{D_2}(g,m),...,A_{D_d}(g,m)$; and in the third method, the likelihood are calculated in the same way as the second method, except that the covariance matrix is estimated by the our bucket-based covariance estimator.





 

\begin{table}[h]
  \caption{Simulation Results: FDR is the false discovery rate which should be bounded by FDR $\leq$ 0.1 in our setting, and Power is the ratio of true rejections to the number of tests where $H_1$ is true.}
  \centering
  \normalsize
  \label{tab:freq}
  \begin{tabular}{lrr}
    \toprule
    Method&FDR&Power\\
    \midrule
    Non-Covariance& 0.179& 0.827\\
    True Corvariance& 0.079& 0.693\\
    Estimated Corvariance ($B=300$)& 0.103& 0.742\\
    Estimated Corvariance ($B=200$)& 0.115& 0.753\\
  \bottomrule
\end{tabular}
\end{table}

As summarized in TABLE IV, the non-covariance method fails to control FDR although it demonstrates the highest power, while the other methods that take covariance into consideration show better FDR results. It is worth notice that FDR is inflated slightly when the number of buckets reduced from 300 to 200, which may be caused by the precision of the covariance matrix estimation, in fact, we should choose a higher number of bucket $B$ when the number of periods in experiments increase, because the number of parameters in the covariance matrix grows as the periods increase. In reality, most experiments will not be continuing more than 30 days, and a bucket number of 300 is sufficient enough in practice, thus a huge improvement of performance in estimating the covariance matrix is achieved when we have tens of million of observations within each period.

\subsection{Bayesian Optimization}

\subsubsection{Background}
Bayesian optimization is a powerful method for optimizing expensive black-box functions \cite{jones1998efficient}. This method commonly starts with evaluating the objective functions at a few randomly-selected points, and then fitting a surrogate model to the collected data. The posterior surrogate model can provide the estimation of the function value as well as its uncertainty at each point, then we can construct an acquisition function that can balance between exploitation and exploration from this posterior distribution to determine the next query points. The optimization process proceeds sequentially several rounds, fitting the surrogate model by all the data collected at each iteration.
In online controlled experiments, Bayesian optimization can solve the problem of searching the optimal values of continuous parameters with online metrics as the objective; online metrics we observe can be noisy, taking online video playtime per-user as an example: let metric $m$ be everyday playtime, and we have observations $X_{t}(g,u,m)$ for users $u \in \{1,2, ..., U\}$ in group $g$ from Day 1 to Day $d$ $\{D_1, D_2, ..., D_d\}$, then we can define an average metric by $A_{t}(g,m)$, and the average metric we observe in each iteration is a normal distributed random variable.

In real-world scenarios, the experimenters would usually want a complex objective composed of multiple online metrics, for example, we may use an objective with the form $af_1 + bf_2$, where $a$ and $b$ are weights, $f_1$ and $f_2$ are two different online metrics, and the two metrics are usually correlated with each other. For example, $f_1$ is the watch time per user and $f_2$ is the number of comments per user. There is a correlation between them because users who watch for a long time are more likely to comment. It is expensive to estimate covariance directly when the sample size reaches tens of millions, so the experimenters would sometimes ignore the effect of correlation between metrics in practice. With the bucket-based covariance estimation method, we can estimate the covariance $\text{cov}(f_1, f_2)$ cost-effectively, thereby improving Bayesian optimization's accuracy and efficiency by taking the correlations of metrics into consideration.

\subsubsection{Numerical Examples}
We compared the optimization performance with covariance estimation and without covariance estimation. The objective function is $g(\mathbf{x}) = 2f_1(\mathbf{x}) + f_2(\mathbf{x})$ where $f_1(\mathbf{x})$ and $f_2(\mathbf{x})$ are the test function Hartmann6 with six dimensions. Hartmann6 has six local minima and one global minimum -9.96711 in $\mathcal{X} = \left\{x_i \in (0, 1), i = 1, ..., 6\right\}$. Our goal is to find the minimum function value at $\mathcal{X}$ which we call best objective here. For each point $\mathbf{x} \in \mathcal{X}$, the true function value is $2f_1(\mathbf{x}) + f_2(\mathbf{x})$, and the observation are computed by $n=10000$  samples from a bivariate normal distribution with mean $\mathbf{\mu} = [2f_1(\mathbf{x}), f_2(\mathbf{x})]$ and covariance matrix $\Sigma = [[\sigma_1^2, \text{cov}(f_2(\mathbf{x}), f_1(\mathbf{x}))]^T, $ $[\text{cov}(f_1(\mathbf{x}), f_2(\mathbf{x})), \sigma_2^2]^T]$. Figure \ref{bo_simulate_results} below shows that covariance estimation can significantly improve Bayesian optimization's convergence speed in both positive and negative correlation.

In the simulation, the covariance matrix for the upper plot is [[0.1375,0.10825318],[0.10825318,0.1125]] which represents $f_1$ and $f_2$ have a positive correlation; and the covariance matrix for lower plot is [[0.084375,-0.11095398],[-0.11095398,0.1700625]] which represents $f_1$ and $f_2$ have a negative correlation. We can see from the results that the procedure with covariance estimation outperformed the procedure without covariance estimation.
\begin{figure}[h]
  \centering
        \begin{minipage}{.5\textwidth}
        \centering
        \includegraphics[scale=0.4]{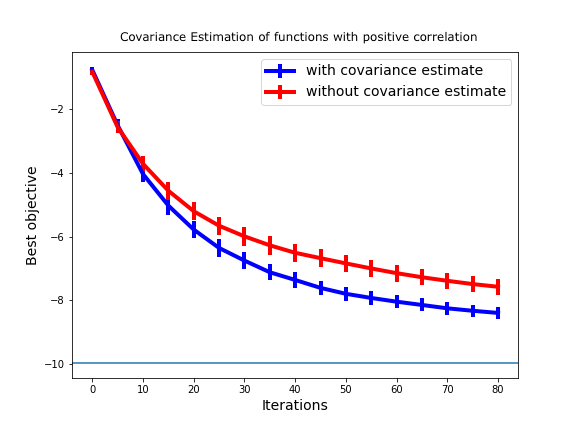}
        \end{minipage}
      \begin{minipage}{.5\textwidth}
      \centering
      \includegraphics[scale=0.4]{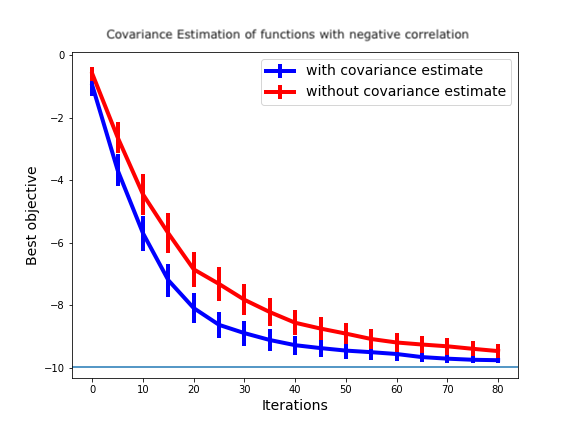}
      \end{minipage}
  \caption{Value of the best objective by each iteration of optimization, with horizontal line indicates the global optimum for the problem, and the lower line is approaching the global optimum faster.}
  \label{bo_simulate_results}
\end{figure}

\section{Benchmarks in Practice}
\label{sec:benchmarks}
We now demonstrate the performance benchmarks for our bucket-based covariance estimator in a subset of our company's running experiments, which contains 345 experiments from one business scenery of our company. For every metric in our system, there are approximately 10 billion samples from these experiments every day, and for every metric we have a corresponding daily routine which divert these samples into a number of $B$ (typically less than 1000) buckets for each experiment group, then the covariance estimation of any two days can be calculated through web services interactively and simultaneously using these buckets reduced from samples. However, for data-join method we have to set up an offline task which can join the two days of data together by users and then estimate the covariance for each experiment group by the data augmentation method described previously. 

Figure 3 displays the benchmarks of estimating the pairwise covariance of one metric within 2,3,4 days respectively. For our bucket-based method we need to preprocess samples from every day respectively, thus the complexity grows linearly as the number of days $n$ increases; while for the data-join method we need to launch an offline task for each pair of days, which lead to a complexity of $O(n^2)$, and it will be difficult to compute when we have thousands of running experiments, with thousands of metrics being calculated and one billion active users participating in our system.

\begin{figure}[h]
  \centering
  \includegraphics[width=\linewidth]{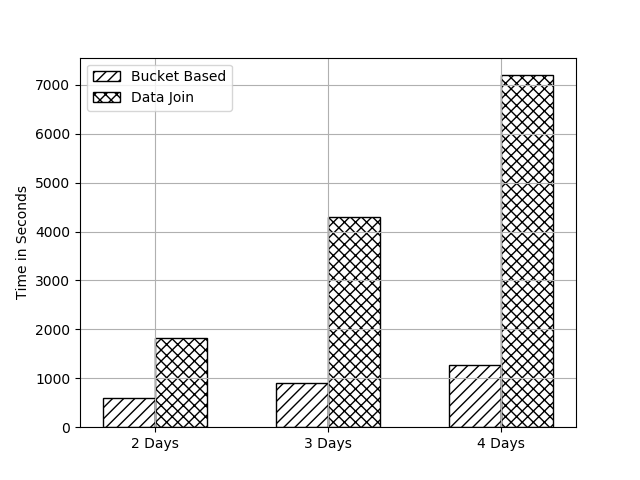}
  \caption{Running time for pairwise covariance estimation of one metric for 345 running experiments among different days, a corresponding spark program which has 300 executors and 4G memory for each executor is employed in each method.}
\end{figure}

\section{Conclusion}
Starting from the classical framework of Rubin casual model and the independence assumptions of the hash functions in overlapping online controlled experiment infrastructure, we propose a bucket-based covariance estimation method and theoretical-rigorously prove that this estimator can be generalized to calculate any covariance among different metrics in different periods. Moreover, this method can be easily deployed in numerous applications in any online controlled experiments platform, and we have illustrated the effectiveness and advantages of introducing our covariance estimation method in these applications by both simulation and real online data. The bucket-based method can be flexible to trade off between computational costs and precision in large-scale applications, and improve the performance of numerous applications in online controlled experiments in aspects of efficiency, accuracy and effectiveness. 
Although our covariance estimation can help tackle large data problems, the experimental design still need to be carefully carried out, for example, to increase the sensitivity of metrics\cite{xie2016improving,budylin2018consistent}. These problems are left for further research. 

\appendix[Proof of the Theorem 1]

Step 1: For metrics $m$ and $m'$ of group $g$ at $t$ and $t'$ respectively, let $O_{t}(g,u,m):=Y_{t}(g,u,m) Z_{t}(g,u,m)$ and $O_{t'}(g,u,m'):=Y_{t'}(g,u,m') Z_{t'}(g,u,m')$, in fact, $O_{t}(g,u,m)$ is a new potential outcome which is the product of two potential outcomes $Y_{t}(g,u,m)$ and $Z_{t}(g,u,m)$, $O_{t'}(g,u,m')$ as well. We first prove that for any $S_t(g,m)=\sum_u{I_g(u)O_{t}(g,u,m)}$ and $S_{t'}(g,m')=\sum_u{I_g(u)O_{t'}(g,u,m')}$, we have
\begin{equation}
\begin{split}
&E[B(1-E[I_g(u)])K(S_t(g,m),S_{t'}(g,m'))]\\
&= \text{cov}[S_t(g,m),S_{t'}(g,m')]
\end{split}
\end{equation}
Proof of (1). By definition we have $$S_t(g,m,b)=\sum_u{I_g(u)I_b(u)O_{t}(g,u,m)}$$ where $S_{t}(g,m,b)$ is the potential outcome sum of bucket $b$ and $I_b(u)$ is the indicator representing if the user $u$ is assigned to bucket $b$. Thus we have 
\begin{equation}
\begin{split}
&E[K(S_{t}(g,m),S_{t'}(g,m'))]=E[\frac{B}{B-1}[\frac{\sum_b{S_{t}(g,m,b)S_{t'}(g,m',b)}}{B} \\
&-\frac{\sum_b{S_{t}(g,m,b)}}{B} \frac{\sum_{b'}{S_{t'}(g,m',b')}}{B}]]\\
&=E[\frac{\sum_b{S_{t}(g,m,b)S_{t'}(g,m',b)}}{B-1}] \\
&- E[\frac{1}{B(B-1)} \sum_{b}\sum_{b'}{S_{t}(g,m,b)S_{t'}(g,m',b')}]\\
&=E[\frac{B\sum_b{S_{t}(g,m,b)S_{t'}(g,m',b)}}{B(B-1)}]\\
&- E[\frac{1}{B(B-1)} \sum_{b}\sum_{b'}{S_{t}(g,m,b)S_{t'}(g,m',b')}]\\
&=E[\frac{\sum_b{S_{t}(g,m,b)S_{t'}(g,m',b)}}{B}]\\ &- E[\frac{1}{B(B-1)} \sum_{b\neq b'}{S_{t}(g,m,b)S_{t'}(g,m',b')}]
\end{split}
\end{equation}
and 
$$
\begin{aligned}
&E[S_{t}(g,m,b)S_{t'}(g,m',b')]\\
&=E[\sum_u{I_g(u)I_b(u)O_{t}(g,u,m)}\sum_u{I_g(u)I_{b'}(u)O_{t'}(g,u,m')}]\\
&=E[\sum_{u}{I^2_g(u)I_b(u)I_{b'}(u)O_{t}(g,u,m)O_{t'}(g,u,m')}\\
&+\sum_{u\neq u'}{I_g(u)I_g(u')I_b(u)I_{b'}(u')O_{t}(g,u,m)O_{t'}(g,u',m')}]\\
&=E[I^2_g(u)I_b(u)I_{b'}(u)]\sum_{u}{O_{t}(g,u,m)O_{t'}(g,u,m')}\\
&+ E[I_g(u)I_g(u')I_b(u)I_{b'}(u')]\sum_{u\neq u'}{O_{t}(g,u,m)O_{t'}(g,u',m')}
\end{aligned}
$$ 
Notice that $I_b(u)I_{b'}(u)$ always equals to 0 when $b\neq b'$, and from the assumption that the hash function with different seeds generate random results independently, and a hash function with a specific seed generate i.i.d. results for different users, we have:
\begin{equation}
\begin{split}
&E[S_{t}(g,m,b)S_{t'}(g,m',b')]\\
&=\frac{E^2[I_g(u)]}{B^2}\sum_{u\neq u'}{O_{t}(g,u,m)O_{t'}(g,u',m')}
\end{split}
\end{equation} if $b\neq b'$, and
\begin{equation}
\begin{split}
&E[S_{t}(g,m,b)S_{t'}(g,m',b')]\\
&=\frac{E[I_g(u)]}{B}\sum_u{O_{t}(g,u,m)O_{t'}(g,u,m')}\\
&+\frac{E^2[I_g(u)]}{B^2}\sum_{u\neq u'}{O_{t}(g,u,m)O_{t'}(g,u',m')}
\end{split}
\end{equation} if $b = b'$. 

Apply (3) and (4) to (2) we have
\begin{equation}
\begin{split}
&E[K(S_{t}(g,m),S_{t'}(g,m'))]\\
&=\frac{E[I_g(u)]}{B}\sum_u{O_{t}(g,u,m)O_{t'}(g,u,m')}
\end{split}
\end{equation} 
By definition we have 
$$
\begin{aligned}
&\text{cov}[S_{t}(g,m),S_{t'}(g,m')]\\
&=\text{cov}[\sum_u{I_g(u)O_{t}(g,u,m)},\sum_u{I_g(u)O_{t'}(g,u,m')}\\ 
&=\sum_u\sum_{u'} \text{cov}[I_g(u)O_{t}(g,u,m),I_g(u)O_{t'}(g,u',m')]\\
&=\sum_u{I_g(u)O_{t}(g,u,m),I_g(u)O_{t'}(g,u,m')]}\\
&=\sum_u{O_{t}(g,u,m)O_{t'}(g,u,m') cov[I_g(u),I_g(u)] }\\
&=E[I_g(u)](1-E[I_g(u)])\sum_u{O_{t}(g,u,m)O_{t'}(g,u,m')}
\end{aligned}
$$
Then compare the result above to (5), the proof is complete.

Step 2: Let $N$ be the total number of potential users in the population, we have $$(\frac{S_t(g,m)}{N},\frac{N_t(g,m)}{N},\frac{S_{t'}(g,m')}{N},\frac{N_{t'}(g,m')}{N}):=(a,b,c,d)$$ which are asymptotically multivariate normal distributed, according to the alternative form of Delta method, we have $$\text{cov}[\frac{a}{b},\frac{c}{d}] \approx \frac{\text{cov}[a,c]}{bd} + \frac{ac\cdot \text{cov}[b,d]}{b^2d^2} - \frac{c\cdot \text{cov}[a,d]}{bd^2} - \frac{a\cdot \text{cov}[b,c]}{b^2d}$$ Proof Done.


%




\ifCLASSOPTIONcaptionsoff
  \newpage
\fi



\bibliographystyle{IEEEtran}
\bibliography{IEEEabrv,IEEEexample}

\end{sloppypar}
\end{document}